# An *O(n)* method of calculating Kendall correlations of spike trains

William Redman *

Dynamical Neuroscience Program, UC Santa Barbara, Santa Barbara, CA, United States of America

* wredman@ucsb.edu

## Abstract

The ability to record from increasingly large numbers of neurons, and the increasing attention being paid to large scale neural network simulations, demands computationally fast algorithms to compute relevant statistical measures. We present an *O(n)* algorithm for calculating the Kendall correlation of spike trains, a correlation measure that is becoming especially recognized as an important tool in neuroscience. We show that our method is around 50 times faster than the *O*(*n* ln *n*) method which is a current standard for quickly computing the Kendall correlation. In addition to providing a faster algorithm, we emphasize the role that taking the specific nature of spike trains had on reducing the run time. We imagine that there are many other useful algorithms that can be even more significantly sped up when taking this into consideration. A MATLAB function executing the method described here has been made freely available on-line.

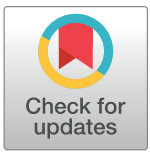

## Introduction

The Kendall correlation was first introduced by Maurice Kendall in 1938 [1]. As a rank correlation, it takes into account the specific ordering of the elements of the sets it is correlating. A Kendall correlation, $\tau$, equal to 1 is interpreted as the elements in the two sets being ordered in the same way. $\tau = -1$ is interpreted as the elements in the two sets being ordered exactly oppositely. And $\tau = 0$ is interpreted as the ordering of the two sets having no relation to one another.

Despite being used in a number of other scientific fields [2–4], it is only recently that the Kendall correlation has started to become appreciated, and implemented, in neuroscience. In particular, due to the usual sparseness of spike trains (i.e. the large number of zeros), the Kendall correlation has been shown to be particularly appropriate for computing pairwise correlations between spike trains, especially as compared to Pearson's correlation [5–7]. Recently, it was used to explore the place field structure of place cells in the hippocampus [7], and generally pairwise correlations can be useful for revealing aspects of the behavior of the recorded, or constructed (in the case of computational/theoretical studies), networks. We note that for the remainder of the paper, by spike train we mean specifically a vector of length $n$ whose i[th] element is a 1 if the corresponding neuron fired at least once during the i[th] time bin of the recorded interval and 0 otherwise. This is a frequently used way to talk about spike trains and is appropriate if firing is particularly sparse or if the time bin size is sufficiently small.

A simple, non-optimized, way of computing the Kendall correlation of two row vectors, $X$ and $Y$, is MATLAB's function, corr(X, Y, 'Type', 'Kendall'). On MATLAB's website [8], they define the Kendall correlation as

$$\tau = \frac{2K}{n(n-1)} \quad (1)$$

where $K = \sum_{i=1}^{n-1} \sum_{j=i+1}^{n} \xi^*(X_i, X_j, Y_i, Y_j)$, and

$$\xi^*(X_i, X_j, Y_i, Y_j) = \begin{cases} 1 & (X_i - X_j)(Y_i - Y_j) > 0 \\ 0 & (X_i - X_j)(Y_i - Y_j) = 0 \\ -1 & (X_i - X_j)(Y_i - Y_j) < 0 \end{cases} \quad (2)$$

However, as additionally stated, MATLAB's function also has a normalization constant in the calculation of $\tau$ that adjusts for ties [8]. A Kendall correlation that takes this additional consideration into account is often referred to as $\tau_b$ in the literature [9]. Therefore, the true way in which MATLAB calculates the Kendall correlation of the row vectors $X$ and $Y$ is

$$\tau = \frac{K}{\sqrt{(n_0 - n_1)}\sqrt{(n_0 - n_2)}} \quad (3)$$

where $n_0 = n(n-1)/2$, $n_1 = \sum_i t_i(t_i - 1)/2$, and $n_2 = \sum_j u_j(u_j - 1)/2$. The sums of $n_1$ and $n_2$ are over all the distinct values $X$ and $Y$ take (respectively), and $t_i$ is the number of elements in $X$ equal to the i$^{th}$ distinct value of $X$ ($u_j$ is the same, but for $Y$).

As can be seen from the definition of $K$, calculating $\tau$ requires summing over many of the pairs of values in $X$ and $Y$ (in fact, $n(n-1)/2$ pairs, which means that the run time is $O(n^2)$). For large spike trains, this results in a large computation time. For this reason, a faster, $O(n \ln n)$ method was developed [10], which makes use of the existence of a mapping between sorting and Kendall correlation. Additional work has been done using sorting and balanced tree structures in cutting edge ways to decrease the run time of other $O(n \ln n)$ methods [11]. While these methods—we will below consider specifically Knight's method [10]—have great power because they are valid for arbitrary vectors, like the $O(n^2)$ method implemented by MATLAB, the generality is unnecessary for computing the Kendall correlation of spike trains. Below, we specifically take the inherent structure of spike trains (that is, that their elements take values only from {0, 1}) under consideration to derive a faster method of calculating Kendall correlations specific to spike trains. We show that our new method is $O(n)$ and then examine how much faster our method is than Knight's method under various conditions.

## Materials and methods

As mentioned above, the motivating idea for the following method is that, since spike trains take values only in {0, 1}, by taking this fact under consideration, we might be able to speed up the calculation of the Kendall correlation. In particular, we show that we can write an explicit formula for $K$ (from Eq (1)) that can be evaluated very quickly—in fact, in $O(n)$.

Considering Eq (2), we see that there are two principle cases we need to consider when calculating $K$: the case where $X_i$ and $X_j$ are in the same order as $Y_i$ and $Y_j$ (i.e. where $\xi^*(X_i, X_j, Y_i, Y_j) = 1$), and the case where they are in the opposite order (i.e. where $\xi^*(X_i, X_j, Y_i, Y_j) = -1$).

The third case, $\xi^*(X_i, X_j, Y_i, Y_j) = 0$, obviously doesn't contribute to the value of $K$. We now consider these two cases separately.

### Same order case

This case happens only when $X_i = Y_i = 1$ and $X_j = Y_j = 0$, or when $X_i = Y_i = 0$ and $X_j = Y_j = 1$ (for $i < j$).

We define the *active set* of $X$ to be

$$A^X = \{i \mid X_i = 1\} \tag{4}$$

where $1 \leq i \leq n$. We similarly define the active set of $Y$, $A^Y$.

We now define the *combined active set*, or the set of positions in the spike trains such that $X_i = Y_i = 1$, as

$$A = A^X \bigcap A^Y = \{i \mid X_i + Y_i = 2\} \tag{5}$$

Now let $N = \{1, 2, \ldots, n\}$. We define the *silent set* of $X$ as

$$S^X = N \setminus A^X \tag{6}$$

where $\cdot \setminus \cdot$ is the set minus operator. We similarly define the silent set of $Y$, $S^Y$.

We now define the *combined silent set*, or the set of positions in the spike trains such that $X_j = Y_j = 0$, as

$$S = S^X \bigcap S^Y = \{j \mid X_j + Y_j = 0\} \tag{7}$$

With Eqs (5) and (7), we can find the contribution to $K$ from this case. The number of ways $\xi^*(X_i, X_j, Y_i, Y_j) = 1$, $K^+$, is

$$K^+ = \sum_{i \in A} |\{j \in S \mid i < j\}| + \sum_{j \in S} |\{i \in A \mid j < i\}| \tag{8}$$

where $|\cdot|$ is the function that returns the number of elements of the set. We see clearly that the first sum in $K^+$ is the number of ways $X_i = Y_i = 1$ and $X_j = Y_j = 0$, and the second sum in $K^+$ is the number of ways $X_i = Y_i = 0$ and $X_j = Y_j = 1$.

By the relationship between the two sums in Eq (8), we can simplify $K^+$ to be

$$K^+ = \sum_{i \in A} |\{j \in S \mid i < j\}| + \sum_{i \in A} (|S| - |\{j \in S \mid i < j\}|) = \sum_{i \in A} |S| = |A| \cdot |S| \tag{9}$$

### Opposite order case

This case happens only when $X_i = Y_j = 1$ and $X_j = Y_i = 0$, or $X_i = Y_j = 0$ and $X_j = Y_i = 1$ (for $i < j$).

We define the *difference of X* as

$$\Delta X = A^X \setminus A^Y = \{i \mid X_i - Y_i = 1\} \tag{10}$$

We similarly define the difference of $Y$, $\Delta Y$. $\Delta X$ is the set of positions in the spike trains where $X_i = 1$ and $Y_i = 0$ (vice versa for $\Delta Y$).

With these we can now find the contribution to $K$ from this case. The number of ways $\xi^*(X_i, X_j, Y_i, Y_j) = -1$, $K^-$, is

$$K^- = \sum_{i \in \Delta X} |\{j \in \Delta Y \mid i < j\}| + \sum_{i \in \Delta Y} |\{j \in \Delta X \mid i < j\}| \tag{11}$$

where the first sum in $K^-$ is the number of pairs $(i, j)$ (where $i < j$) such that $X_i = Y_j = 1$ and $X_j = Y_i = 0$, and the second sum in $K^-$ is the number of pairs $(i, j)$ such that $X_i = Y_j = 0$ and $X_j = Y_i = 1$.

Again, the sums are related (as they were in Eq (8)), so we can re-write $K^-$ as

$$K^- = \sum_{i \in \Delta X} |\{j \in \Delta Y \mid i < j\}| + \sum_{i \in \Delta X} (|\Delta Y| - |\{j \in \Delta Y \mid i < j\}|) = |\Delta X| \cdot |\Delta Y| \quad (12)$$

### Ties

The final thing needed in order to calculate $K$ is the number of tied pairs in $X$ and $Y$, $n_1$ and $n_2$. This is easy in the case of spike trains, as the number of ties for the value 1 is just the sum of all the elements in the train, and the number of ties for the value 0 is just $n$ minus that sum. Therefore, using the equation given for $n_1$, we have

$$n_1 = \frac{1}{2}\left(\sum_{i=1}^{n} X_i (\sum_{i=1}^{n} X_i - 1) + (n - \sum_{i=1}^{n} X_i)(n - \sum_{i=1}^{n} X_i - 1)\right) \quad (13)$$

The same is true for $n_2$ (with $Y$ in place of $X$).

Therefore, with Eqs (9), (12) and (13), we can write the Kendall correlation, Eq (3), of two neural spike trains as

$$\tau = \frac{K^+ - K^-}{\sqrt{(n_0 - n_1)}\sqrt{(n_0 - n_2)}} \quad (14)$$

where $K^+$, $K^-$, $n_0$, $n_1$, and $n_2$ can be found with the formulas we have given for them. Note that Eqs (5), (7), (9), (10), (12) and (13) are all linear in $n$, i.e. $O(n)$. Therefore, Eq (14) is $O(n)$.

### Comparison

To compare the presented method, Eq (14), with Knight's method and MATLAB's method, we created random binary vectors with a specified "sparseness". Here sparseness refers to the expected fraction of 1s present in the vectors (or, in the neural context, the expected activity over a given time interval). We generated these vectors by using MATLAB's rand function, with which we generated $1 \times n$ vectors with elements uniformly drawn from $(0, 1)$ [12]. We then set every element in each vector that had a value less than the sparseness we specified to 1, and all other elements to 0. Put another way, if $X^{\text{rand}}$ was our random $1 \times n$ vector with elements drawn from $(0, 1)$, then we used the transform

$$\hat{X}_i^{\text{rand}} = \begin{cases} 1 & X_i^{\text{rand}} < \text{sparseness} \\ 0 & \text{otherwise} \end{cases} \quad (15)$$

We then used MATLAB's method, Knight's method, and our method to calculate the Kendall correlation of $\hat{X}^{\text{rand}}$ and $\hat{Y}^{\text{rand}}$ (where $\hat{Y}^{\text{rand}}$ was similarly generated). To record the time it took for each method, we used MATLAB's built-in tic toc function [13]. We did all of the calculations on a 2014 MacBook Air (1.4 GHz Intel Core i5) running MATLAB 2015a.

For details of how we implemented Knight's method, see the S1 Text.

### Results

The results of comparing our method to Knight's and MATLAB's methods, are shown in Fig 1. Unsurprisingly, both our method and Knight's method show considerable advantage over

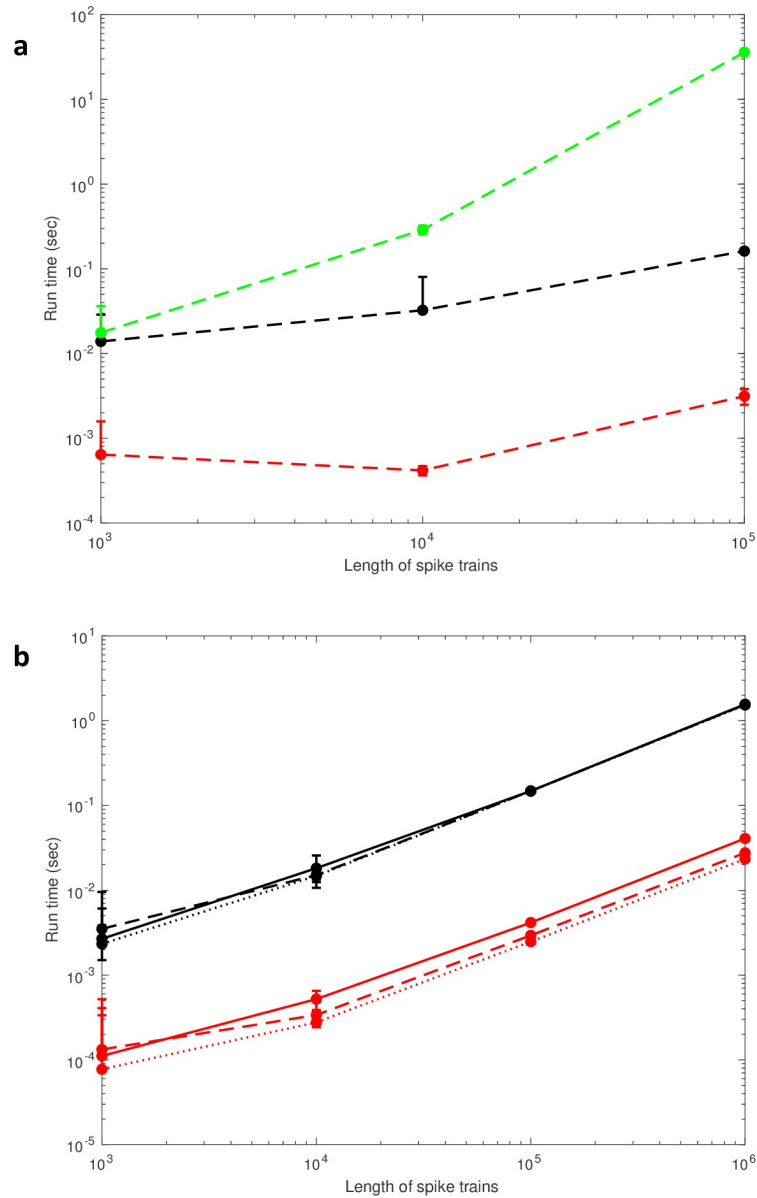

**Fig 1. Run times for all three methods.** (a) The run time as a function of spike train length using Knight's method (black), our method (red), and the standard MATLAB method (green) for a sparseness of 5%. $N = 10$ and error bars are standard deviation. (b) The run time as a function of spike train length for different sparseness values: dotted line (1%), dashed line (5%), solid line (25%). $N = 100$, error bars are standard deviation, and colors are the same as in (a).

https://doi.org/10.1371/journal.pone.0212190.g001

the $O(n^2)$ method that is implemented by MATLAB [8] (Fig 1a). However, our method is definitively faster. Importantly, this holds true for a range of sparseness values (Fig 1b), although our method shows a slight slowing down for larger sparseness values, while Knight's method does not. Our method is on average $\approx 35$ times faster for a sparseness of 25% and $\approx 60$ times faster for a sparseness of 1%. Because a sparseness of 25%, the maximum we tested, is unrealistic for any neural simulation or recording, our method is faster than Knight's in a neurally plausible regime.

**Table 1. Examples of calculated Kendall correlation for all three methods.** Kendall correlation of the spike trains listed at the top of the table (both with length $10^4$) for the three methods.

| | (1010...), (110110...) | (011011...), (110110...) |
|---|---|---|
| $O(n)$ Method | $-2.121373383860751 \times 10^{-4}$ | $-0.500037496719090$ |
| Knight's Method | $-2.121373383860751 \times 10^{-4}$ | $-0.500037496719090$ |
| MATLAB's Method | $-2.121373383860751 \times 10^{-4}$ | $-0.500037496719090$ |

https://doi.org/10.1371/journal.pone.0212190.t001

Finally, for all the correlations between spike trains we computed, we checked that the two Kendall correlation values were within $10^{-12}$ of MATLAB's Kendall correlation function (see Table 1). Therefore, we feel confident that our method is correct and equivalent (up to machine error) to MATLAB's method.

## Discussion

We have presented a novel method to calculate Kendall correlations of large spike trains, and have demonstrated its advantage (in terms of computation time) to the standard for fast Kendall correlation computation [10]. We achieved this by specifically taking the structure of spike trains (the fact that they are made up of 1s and 0s) into consideration, and deriving explicit formulas for the components of the Kendall correlation (Eqs (9), (12) and (13)). These formulas are all linear in $n$, meaning our method is $O(n)$, unlike Knight's method which is $O(n \ln n)$. We have also, by way of computation, provided evidence that our method is correct and equivalent (up to machine error) to MATLAB's standard method.

With a significantly faster method to compute the Kendall correlation between large spike trains, we hope that the Kendall correlation will become a more accessible tool for neuroscience. While we know there are faster ways to implement algorithms similar to Knight's (as was explored in [11]) that may be faster than the method provided here, the simplicity of our method (a few linear equations) makes it much more appealing to neuroscientists who have limited technical knowledge and/or interest in computer science. We imagine it will be especially useful in computational/theoretical studies where large, sparse spike trains are frequently generated and whose pairwise correlations provide insight into the complex properties of the network. We hope that the fact that pairwise correlations over significantly longer time intervals (or equivalently, between spikes trains of longer lengths) can now be calculated quickly, more in-depth analysis of generated networks (in addition to analysis of observed/recorded networks) will be achieved.

Finally, we hope that our results make clear the usefulness of considering specifically the structure of spike trains when calculating certain quantities. We're sure many other measures can be significantly sped up when taking this into consideration.

## Supporting information

**S1 Code.**
(PDF)

**S1 Text. Implementation of Knight's method.**
(PDF)

## Acknowledgments

We thank Eliott Levy for fruitful discussion and mentorship. We thank the reviewers for their helpful and constructive comments that pushed us towards making our algorithm $O(n)$. We dedicate this paper to Prof. David Cai, who was among the first to inspire us towards research in neural science.

## References

1. Kendall M. A New Measure of Rank Correlation. Biometrika 1938; 30 (1–2): 81–89. https://doi.org/10.2307/2332226
2. Slamon DJ, Godolphin W, Jones LA, Holt JA, Wong SG, Keith DE, et al. Studies of the HER-2/neu proto-oncogene in human breast and ovarian cancer. Science 1989; 244: 707–712. https://doi.org/10.1126/science.2470152 PMID: 2470152
3. Giraudet AL, Al Ghulzan A, Auperin A, Leboulleux S, Chehboun A, Troalen F, et al. Progression of medullary thyroid carcinoma: assessment with calcitonin and carcinoembryonic antigen doubling times. Eur J Endocrinol 2008; 158:239–246. https://doi.org/10.1530/EJE-07-0667
4. Kelder T, Stroeve JH, Bijlsma S, Radonjic M, Roeselers G. Correlation network analysis reveals relationships between diet-induced changes in human gut microbiota and metabolic health. Nutrition & Diabetes 2014 Jun 30; 4:e122. https://doi.org/10.1038/nutd.2014.18
5. Press WH, Teukolsky SA, Vetterling WT, Flannery BP. Numerical Recipes: The Art of Scientific Computing. Cambridge: Cambridge Univ. Press 2007.
6. Soletta JH, Farfán FD, Felice CJ. Measuring Spike Train Correlation with Non-Parametric Statistics Coefficient. IEEE Latin America Transactions 2015 Dec https://doi.org/10.1109/TLA.2015.7404902
7. Neymotin SA, Talbot ZN, Jung JQ, Fenton AA, Lytton WW. Tracking recurrence of correlation structure in neuronal recordings. J. Neurosci. Methods 2017; 275: 1–9. https://doi.org/10.1016/j.jneumeth.2016.10.009 PMID: 27746231
8. https://www.mathworks.com/help/stats/corr.html
9. Agresti A. Analysis of Ordinal Categorical Data. 2nd ed. New York: John Wiley & Sons. 2010. ISBN 978-0-470-08289-8. https://doi.org/10.1002/9780470594001
10. Knight WR. A computer method for calculating Kendall's tau with ungrouped data. J. Am. Stat. Assoc. 1966; 61 (314): 436–439. https://doi.org/10.1080/01621459.1966.10480879
11. Christensen D. Fast algorithms for the calculation of Kendall's $\tau$. Computational Statistics 2005; 20: 51–62 https://doi.org/10.1007/BF02736122
12. https://www.mathworks.com/help/matlab/ref/rand.html
13. https://www.mathworks.com/help/matlab/ref/tic.html